\documentstyle[12pt,epsf]{article}

\begin{document}

\begin{center}
{\bf SEARCH FOR $\eta$-MESIC NUCLEI
 \\  IN PHOTO-MESONIC PROCESSES}

\vskip 5mm
G.A.~Sokol$^\dag$,
T.A.~Aibergenov,
A.V.~Kravtsov,
Yu.I.~Krutov,
A.I.~L'vov,
L.N.~Pavlyuchenko,
S.S.~Sidorin
\vskip 5mm

{\small \it
P.N. Lebedev Physical Institute of the Russian Academy of Sciences \\
Leninsky Prospect 53, Moscow 117924, Russia
\\ $\dag$ \it  E-mail: gsokol@sgi.lpi.msk.su}

\vskip 5mm
\begin{minipage}{150mm}
\centerline{\bf Abstract}
\medskip
Results of an experiment performed at the Lebedev Physical
Institute with the bremsstrahlung photon beams of the end-point
energies 650 and 850 MeV are presented.  Correlated $\pi^+n$
pairs with the opening angle $\langle\theta_{\pi N}\rangle =
180^\circ$ and the energies $\langle E_{\pi^+}\rangle=300$ MeV
and $\langle E_n\rangle=100$ MeV from the process $\gamma +
{}^{12}\mbox{C} \to N + {}_\eta(A-1) \to N + \pi^+ n + (A-2)$
were observed.  They provide an evidence for the existence of
the ${}^{11}_{~\eta}$B and ${}^{11}_{~\eta}$C $\eta$-mesic
nuclei.


\end{minipage}
\end{center}

\vskip 10mm

\section{Introduction}

Eta-mesic nuclei ${}_\eta A$ are a new form of the nuclear
matter which represents bound states of the $\eta$-meson and a
nucleus. Till now, the eta-mesic nuclei were not discovered
despite theoretical works of more than 10 years old which
suggested an existence of such exotic objects of the nuclear
physics \cite{hai86,liu86}. Two attempts to discover the
$\eta$-nuclei were performed soon after the first theoretical
suggestions. They were based on using $\pi^+$-meson beams at BNL
\cite{chr88} and LAMPF \cite{lei88}. Both of these experiments
failed to find $\eta$-nuclei, what is probably related with
non-optimal experimental conditions and not quite adequate
interpretation of experimental data.

The negative result of the first experiments cooled
significantly an interest in solving the $\eta$-nuclei problem
and conducting direct searches for $\eta$-nuclei. Meanwhile,
experimental studies of the reactions $d(p,{}^3\mbox{He})\eta$
\cite{ber88} and $^{18}\mbox{O}(\pi^+,\pi^-){}^{18}\mbox{Ne}$
\cite{joh93} did provide an evidence for formation of a
strongly-bound state of the $\eta$-meson and the nucleus in the
intermediate stage of the reactions \cite{kon94}. Moreover,
recent re-evaluations \cite{gre97} of the $s$-wave scattering
length of $\eta N$ scattering, $a_{\eta N}$, give a
significantly larger (by $\sim 3$ times) value of Re\,$a_{\eta
N}$ which makes feasible to exist $\eta$-nuclei at all $A \ge 3$
and, perhaps, even at $A=2$ \cite{rak96}.

These later experimental and theoretical works stimulated new
searches for $\eta$-mesic nuclei. Among new proposals is an idea
to search for $\eta$-mesic nuclei in photo-mesonic reactions
from nuclei \cite{sok94} which is based on a method of
identifying $\eta$-nuclei suggested in Ref. \cite{sok91}. In the
present work, given are the first (preliminary) data of the
experiment on a search for $\eta$-nuclei in photo-reactions with
the bremsstrahlung photon beam of the 1-GeV electron synchrotron
of the Lebedev Physical Institute.  They provide an evidence for
an existence of $\eta$-mesic nuclei.

\section{Search for $\eta$-nuclei in photo-mesonic reactions}

The use of photons rather than pions to produce $\eta$-nuclei
may provide some advantages. Photon beams are far more intense,
and this quite compensates a lower cross section of $\eta$
production in electromagnetic vs. strong interactions. Also,
photons freely penetrate into the nucleus and make all nucleons
to participate in the $\eta$-nucleus production.

The process of the $\eta$-nucleus formation in a
photo-reaction followed by a decay is depictured in Fig.~1. It
is assumed that both the first stage of the reaction, i.e.
production of the $\eta$ by the photon, and the second stage,
i.e. annihilation of the $\eta$ and creating the pion, proceeds
through single-nucleon interactions (either with a proton or a
neutron in the nucleus) mediated by the $S_{11}(1535)$ nucleon
resonance.  Formation of the bound state of the $\eta$ and the
nucleus becomes possible when the momentum of the produced
$\eta$ is small (typically less than 150 MeV/c). Note that the
$\eta N$ interaction is attractive when the kinetic energy of
the free $\eta$ is $T_\eta \le 50$ MeV \cite{gre97}. These
restrictions suggest the energies of photons, $E_\gamma =
650{-}850$ MeV, as most suitable for creating $\eta$-nuclei.

\begin{figure}[h]
\leavevmode
\epsfxsize=0.45\textwidth\epsfbox[48 653 465 800]{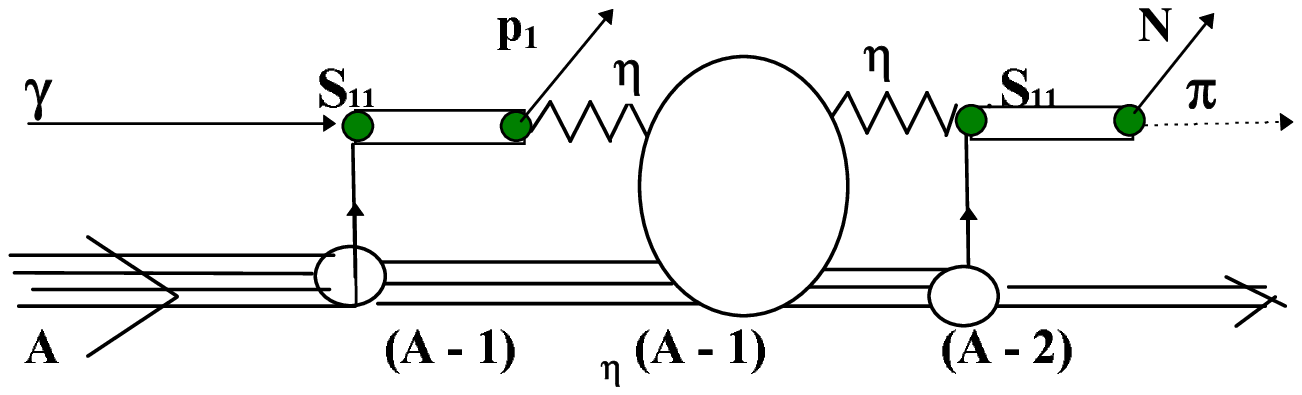}
\leavevmode \hfill
\epsfxsize=0.45\textwidth\epsfbox[48 653 465 800]{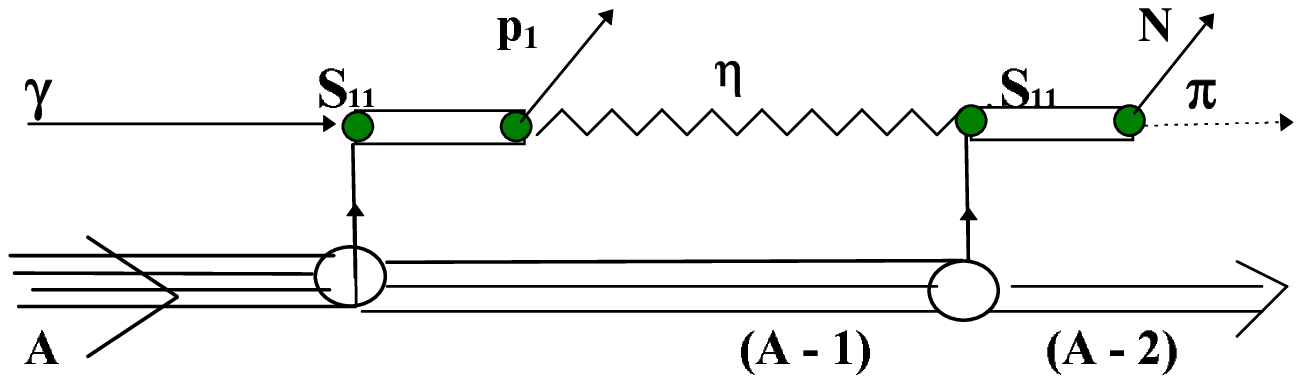}
\\
\parbox[b]{0.45\textwidth}
{\caption{Mechanism of formation and decay of an $\eta$-nucleus.}
\label{fig1}}
\hfill
\parbox[b]{0.45\textwidth}
{\caption{Background production and decay of $\eta$'s in the nucleus.}
\label{fig2}}
\end{figure}

In cases when the momentum (or energy) of the produced $\eta$ is
high, the attraction between the $\eta$ and the nucleus is not
essential, and the $\eta$ propagates freely, see Fig.~2. In
this case the final $\pi N$ pairs carry a high momentum too and
their kinematical characteristics, such as an opening angle, are
different from those of pairs produced through the stage of the
$\eta$-nucleus formation.

\section{Identification of $\eta$-nuclei}

In the present experiment, detection of the $\pi N$ pairs from
the decay of the second-stage $S_{11}$ resonance was selected as
a trigger \cite{sok91}. Such pairs have an (almost) isotropic
distribution.  When the energy and the momentum of the
intermediate $\eta$ are low (as in the case of the mechanism
shown in Fig.~1), the characteristic opening angle of the pair
is $\langle\theta_{\pi N}\rangle = 180^\circ$ with the width of
$\simeq 25^\circ$ due to Fermi motion, and the kinetic energies
are distributed around $\langle E_\pi\rangle \simeq 300$ MeV and
$\langle E_n\rangle \simeq 100$ MeV.  In the case of a higher
momentum of $\eta$ (diagram in Fig.~2), the opening angle is
smaller.

Thus, identification of $\eta$-nuclei can be realized by
analyzing angular and energy distributions of the $\pi N$ pairs.
Spectrum of the total energy of the pairs is expected to form a
peak at $E_{\rm tot} \simeq 1460$ MeV of the total width 25--50
MeV \cite{lvo98}. The reduction of the total energy of the
pairs vs. $m_\eta+m_N=1486$ MeV is related with formation of the
quasi-bound $\eta$-nucleus state(s) and with an excitation of
the rest of the nucleus.  Analysis of spectra observed in the
actual experiment was done with taking into account energy
losses in absorbers and detectors.

\section{Performance of the experiment}

The experiment was performed at the bremsstrahlung photon beam
of the electron synchrotron ``Pakhra" of the Lebedev Physical
Institute.  Parameters of the beam are: $E_{e\,\rm max}=1.2$
GeV, $f=50$ Hz, $I_e = 10^{12}\,s^{-1}$, $\Delta\tau_\gamma=2$ ms
(duration of the photon beam bunch), the duty factor $= 0.1$.

The experimental setup consisted of two scintillation
time-of-flight spectrometers with the apparatus time resolution
(the channel acceptance) of $\Delta\tau=50$ ps. Both
spectrometers were positioned around a 4 cm carbon target at
either $\theta=50^\circ$ or $90^\circ$ with respect to the
photon beam (on its opposite sides). Among four possible charge
combinations of $\pi N$ pairs produced from $\eta$-meson
collisions inside the nucleus, i.e.  $\pi^+n$, $\pi^-p$,
$\pi^0p$ and $\pi^0n$, the first one was chosen as the most
suitable for effective detection and identification.  In Fig.~3,
a layout of the $\pi^+$ and $n$ spectrometers is shown with
typical time-of-flight spectra for $\pi^+n$ coincidences. The
spectrometers were build of scintillator blocks of
$500\times500\times100$ mm$^3$ and $500\times500\times20$ mm$^3$
with four photo-tubes FEU-63 and FEU-143 located at corners of
the scintillator plates. Coordinates of particles passed through
the detectors were determined via a time difference of the light
collection by the photo-tubes. This procedure gave a space
resolution $\sigma_x=\sigma_y \simeq 5$ mm. The time-of-flight
base between the first ($T1$) and last ($T2$) detectors of the
pion spectrometer was 1~m. The start signal was formed by the
$T1$ detector.  The trigger was a coincidence of two signals
from the $\pi$ spectrometer ($T1\land T2$) with a signal from
any detector of the $n$ spectrometer ($N_i$) and an
anti-coincidence with the anti-counter $A$ of charged particles
located in front of the neutron detectors and having the
efficiency of 90\%.

\begin{figure}
\leavevmode
\epsfxsize=0.35\textwidth\epsfbox[114 560 305 769]{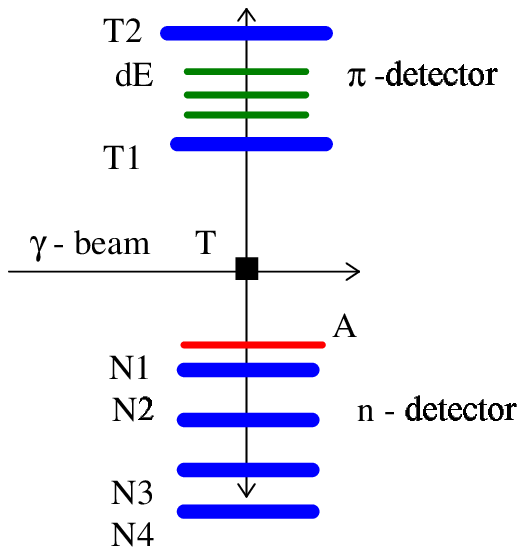}
\leavevmode
\epsfxsize=0.30\textwidth\epsfbox[000 000 184 220]{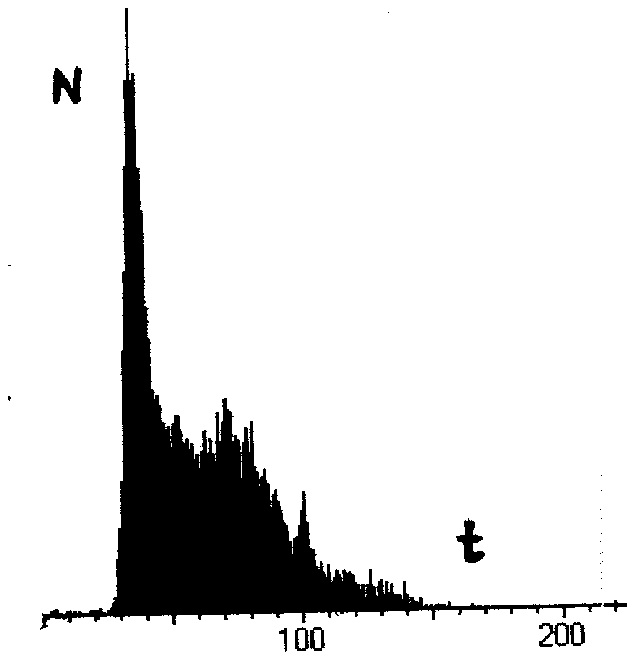}
\leavevmode
\epsfxsize=0.30\textwidth\epsfbox[000 000 200 210]{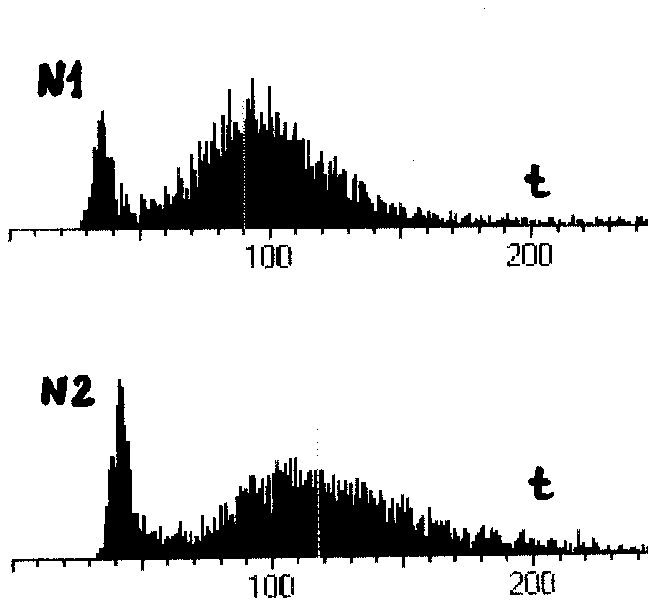}
\vspace{-2em}
\caption{Layout of the experimental setup. Also shown are
time-of-flight spectra in the $\pi$ (left) and $n$
(right) detectors.}
\label{fig3}
\end{figure}

There were three runs in the present experiment with different
positions of the spectrometers: a) ``calibration", b)
``background", and c) ``effect $+$ background". In the
``calibration" run a), both spectrometers were placed at
$\theta=50^\circ$ with respect to the photon beam, and the
end-point energy of the bremsstrahlung spectrum was $E_{\gamma
\rm max}=650$ MeV.  In this run, mainly $\pi^+n$ pairs from
quasi-free production of pions from the carbon were detected:
\begin{equation}
\label{QF}
   \gamma + {}^{12}\mbox{C} \to \pi^+ +  n + X.
\end{equation}
Since energy characteristics of pions and neutrons in the
reaction (\ref{QF}) at $\theta=50^\circ$ are similar to those
expected for the decay products of $\eta$'s in the nucleus, this
run was used to calibrate the setup and to adjust time delays
for the spectrometers.

In the ''background" run b), the spectrometers were moved at
$\theta=90^\circ$ with respect to the photon beam, i.e. to the
position suitable for measuring the effect. However, the
end-point energy still was $E_{\gamma \rm max}=650$ MeV, i.e.
well below the $\eta$ photoproduction threshold off free
nucleons (707 MeV), so that $\eta$'s were not produced.

In the last run c), keeping the angle $\theta=90^\circ$, the
beam energy was increased up to $E_{\gamma \rm max}=850$ MeV,
i.e. above the $\eta$ threshold.  Then a large excess of the
correlated $\pi^+n$ pairs was experimentally observed.

\section{Data analysis}

The observed two-dimensional velocity spectra of the detected
pairs are shown in Fig. 4 for all three runs.  In accordance
with the velocities, all events in each run can be assembled
into three groups:  fast-fast (FF), fast-slow (FS), and
slow-slow (SS).

The FF events with the extreme velocities close to the speed of
the light correspond to a background (mainly $e^+e^-$ pairs
produced by $\pi^0$ from double-pion production).

The FS events mostly correspond to $\pi N$ pairs.

In the ``calibration" run ($\theta=50^\circ$, $E_{\gamma \rm
max}=650$), the quasi-free production of the $\pi^+n$ pairs is
seen as a prominent peak in the two-dimensional distribution
(see Fig.~4a).

In the ``background" run ($\theta=90^\circ$, $E_{\gamma \rm
max}=650$), the largest peak (SS events in Fig.~4b) is caused
by $\pi\pi$ pairs from double-pion photoproduction off the
nucleus.

In the ``effect$+$background" run ($\theta=90^\circ$ and
$E_{\gamma \rm max}=850$ MeV), apart from the SS events, a clear
excess of the FS events as compared with the ``background" run
is seen.  This FS signal is interpreted as a result of
production and decay of slow etas in the nucleus giving the
$\pi^+n$ pairs.

In order to reduce a mis-identification of a $e^+e^-$ pair as a
single pion, a further analysis of the events was done by using
information from three scintillation detectors which were
positioned between $T1$ and $T2$ and measuring the energy losses
$\Delta E$.  The two-dimensional distribution over the
time-of-flight $T$ and the energy loss $\Delta E$ (see Fig.~5)
allows to discriminate the events with a single pion from the
$e^+e^-$ background by selecting events with minimal $\Delta
E$.

The count rate of the $\pi^+n$ events was evaluated as
$$
  N(\pi^+n; 850) =
  N(\mbox{FS}; 850) - N(\mbox{FS}; 650) \times K(850/650),
$$
where $N(\mbox{FS};E_{\gamma \rm max})$ was a number of the
observed FS events at the specified energy $E_{\gamma \rm max}$
and the coefficient K(850/650) was an increase of the
FS-background due to the double-pion photoproduction when
$E_{\gamma \rm max}$ was growing from 650 MeV up to 850 MeV.
This coefficient was determined by using the SS events at 650
and 850 MeV. It was equal to $K=2.15$. This procedure gave
$N(\pi^+n; 850) = (61 \pm 7)$ events/hour.

The differential cross section of producing
the correlated $\pi^+n$ pairs in the reaction (\ref{QF})
in the energy range $\Delta E_\gamma = 650{-}850$ MeV is found as
$$
  \frac{d\sigma}{d\Omega}(\pi^+n; \Delta E_\gamma) =
  \frac{N(\pi^+n)}{N_\gamma N_{\rm nucl}
  \Delta\Omega_\pi \xi_\pi \xi_n f} \, ,
$$
where $N(\pi^+n)=(61\pm 7)$ hour$^{-1}$ is the number of events,
$N_\gamma = 0.75\cdot 10^{11}$ hour$^{-1}$ is
the photon flux in the energy interval $\Delta E_\gamma$,
$N_{\rm nucl}=3.4\cdot 10^{23}$ cm$^{-2}$ is the nuclear density
of the target, $\Delta\Omega_\pi=5.8\cdot 10^{-2}$ sr
is the solid angle of the $\pi$-telescope, $\xi_\pi =  0.8$ is
the detection efficiency of the pions, $\xi_n = 0.3$ is the
detection efficiency of the neutrons (by any of the four neutron
detectors), and $f=0.18$ is the geometrical fraction of the
correlated $\pi^+n$ pairs simultaneously detected by the pion
and neutron detectors.  The last number is determined by
smearing the angular correlations between the pion and neutron
momenta due to Fermi motion and was determined by a Monte Carlo
simulation.

Using these number, we get the differential cross section
$$
  \frac{d\sigma}{d\Omega}(\pi^+n; \Delta E_\gamma) =
  (0.97 \pm 0.11) ~\mu\mbox{b}/\mbox{sr}
$$
and, assuming an isotropic distribution of the pairs,
the total cross section
\begin{equation}
\label{cstot}
  \sigma(\pi^+n; \Delta E_\gamma) =
  4\pi \frac{d\sigma}{d\Omega}(\pi^+n; \Delta E_\gamma) =
  (12.2 \pm 1.3) ~\mu\mbox{b}.
\end{equation}

\begin{figure}
\raisebox{3cm}{a)}
\epsfxsize=10cm\epsfbox[46 492 565 685]{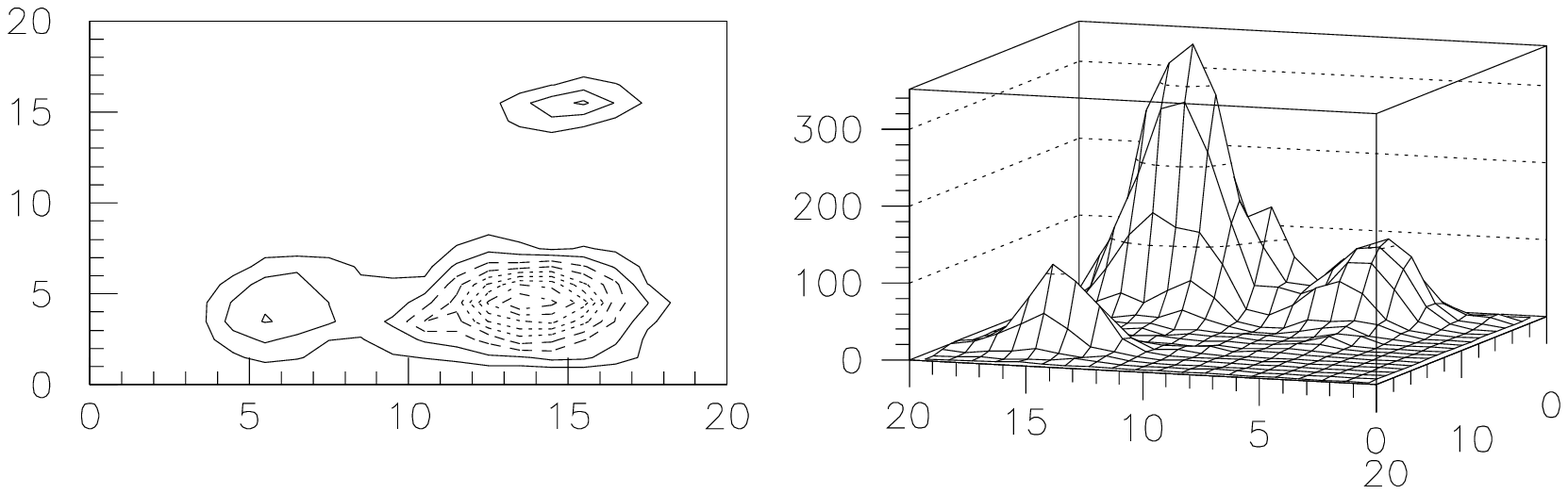}
\hspace{2em}
\raisebox{3.5cm}{a)} \hspace{-3em}
\epsfxsize=5cm\epsfbox[189 332 384 489]{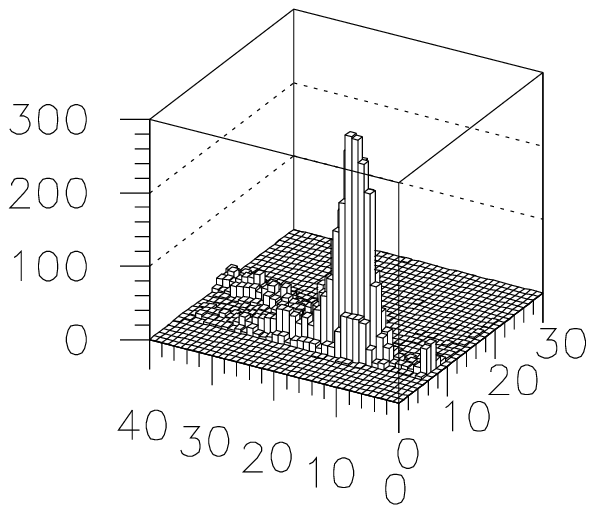}
\\[-1ex]
\raisebox{3cm}{b)}
\epsfxsize=10cm\epsfbox[46 492 565 685]{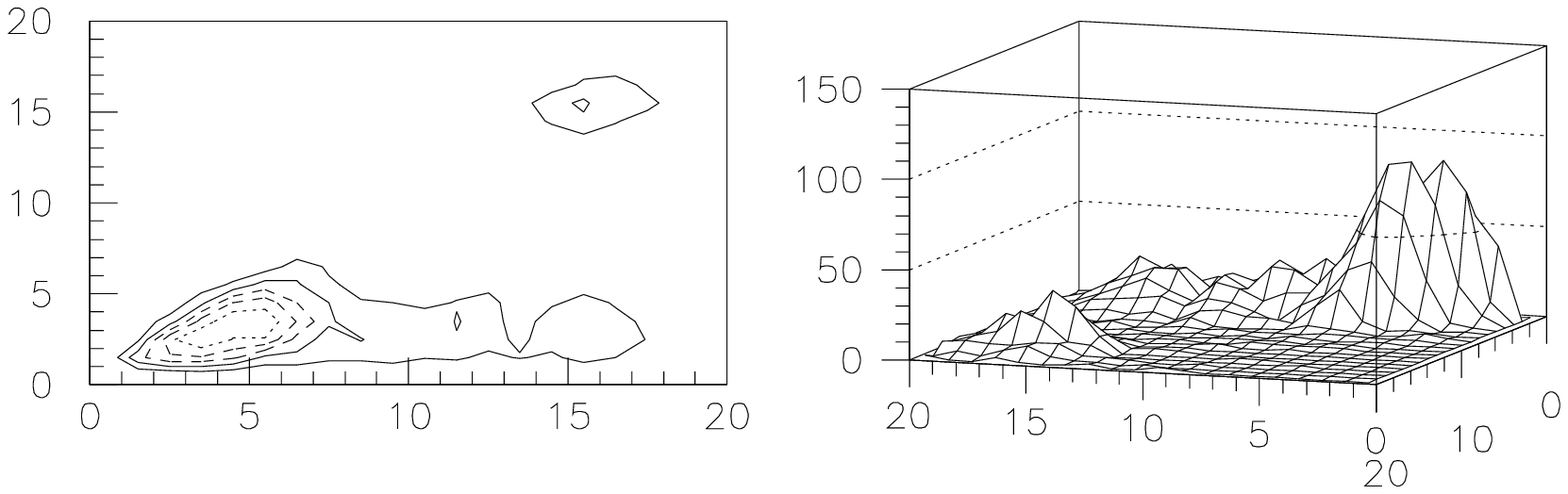}
\hspace{2em}
\raisebox{3.5cm}{b)} \hspace{-3em}
\epsfxsize=5cm\epsfbox[189 332 384 489]{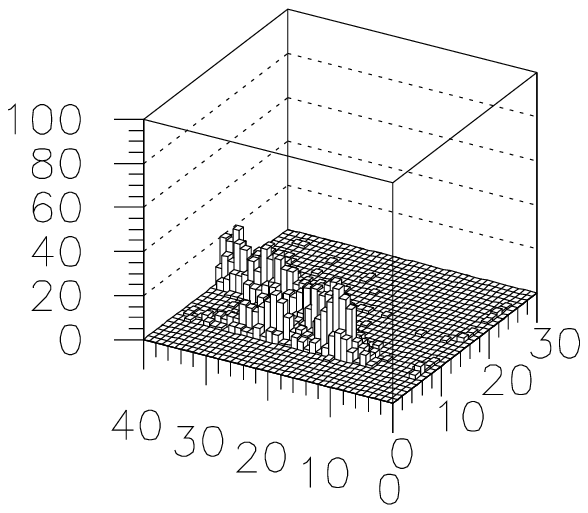}
\\[-1ex]
\raisebox{3cm}{c)}
\hspace{-0.5cm}\raisebox{1.5cm}{$v_n$}
\hspace{2cm}\raisebox{-0.0cm}{$v_\pi$}
\hspace{4cm}\raisebox{-0.0cm}{$v_\pi$}
\hspace{2cm}\raisebox{0.1cm}{$v_n$}
\hspace{-9.6cm}
\epsfxsize=10cm\epsfbox[46 492 565 685]{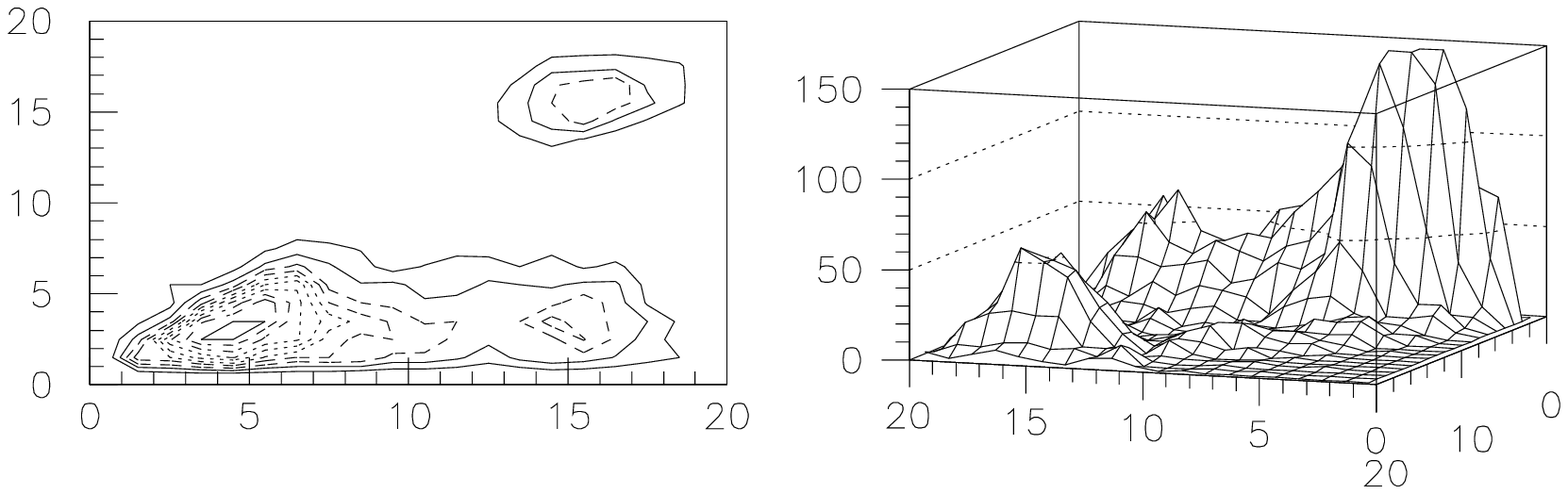}
\hspace{2em}
\raisebox{3.5cm}{c)}
\hspace{0cm}\raisebox{-0.2cm}{$\Delta E$}
\hspace{2.2cm}\raisebox{0.2cm}{$T$}
\hspace{-4.7cm}
\epsfxsize=5cm\epsfbox[189 332 384 489]{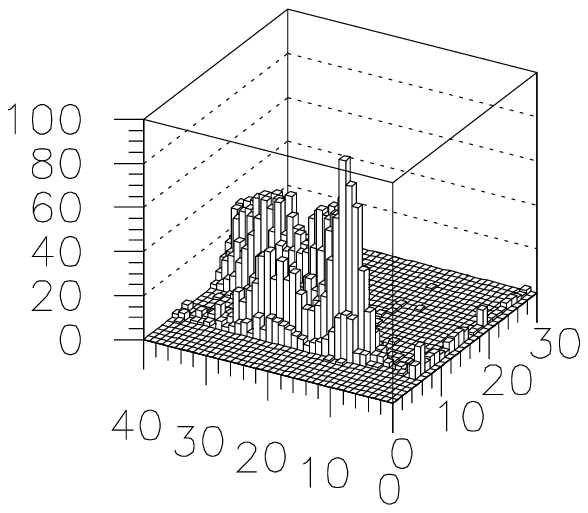}
\\[-1ex]
\parbox[t]{7cm}
{\caption{Distributions over the pion and neutron velocities
 (arbitrary units) for the ``calibration" (a), ``background" (b), and
``effect$+$background" (c) runs.}
\label{fig4}}
\hfill
\parbox[t]{8cm}
{\caption{Distributions over the time-of-flight $T$ and the
energy losses $\Delta E$ in the pion spectrometer (arbitrary
units) for the ``calibration", ``background", and
``effect$+$background" runs. Only events with a slow particle in
the neutron spectrometer were selected.}
\label{fig5}}
\end{figure}

\section{Conclusions}

The observation of the excess of the correlated $\pi^+n$ pairs
when the photon energy becomes higher than the $\eta$-meson
production threshold provides an evidence for production and
decay of $\eta$'s inside the nucleus. The obtained total
photoproduction cross section for such pairs (\ref{cstot}) is
close to theoretical predictions \cite{try95} for the total
cross section of formation of $\eta$-nuclei. This provides a
confirmation that the observed events are related with the
formation of $\eta$-nuclei.  For more direct arguments, angular
and energy distributions of the components of the pairs have to
be analyzed. This work is in progress now.

\section*{Acknowledgements}

The authors thank S.I. Nikolsky and E.I. Tamm for their
attention to and support for the work. A.I. Lebedev and V.A.
Tryasuchev are highly appreciated for valuable discussions of
the physics of $\eta$-nuclei. Special thanks are to G.G.
Subbotin and the synchrotron stuff for the skillful operation
with the synchrotron, to L.I. Goryacheva for her help in
preparing the manuscript, and to N.N. Lvova for her help in
software problems.

This work was supported by the Russian Foundation for Basic
Research, grant 96-02-17103.


\begin{thebibliography}{xx}

\bibitem{hai86} Q. Haider, L. Lui,
   Phys. Lett. {\bf B172}, 257 (1986).

\bibitem{liu86} L. Lui, Q. Haider,
   Phys. Rev. {\bf C34}, 1845 (1986).

\bibitem{chr88} R. Chrien {\it et al.},
   Phys. Rev. Lett. {\bf 60}, 2595 (1988).

\bibitem{lei88} B. Leib, L. Liu,
   LAMPF Progress Report, LA-11670-PR (1988).

\bibitem{ber88} J. Berger {\it et al.},
   Phys. Rev. Lett. {\bf 61}, 919 (1988).

\bibitem{joh93} J. Johnson {\it et al.},
  Phys. Rev. {\bf C47}, 2571 (1993).

\bibitem{kon94} L. Kondratyuk {\it et al.}, Proc. of Intern.
Conf.  "Mesons and Nuclei at Intermediate Energies" May 3-7,
Dubna, Russia 1994, Eds. M.~Khankhasaev, Z.~Kurmanov. Singapore
1995, p.714.

\bibitem{gre97} A.M. Green, S. Wycech,
  Phys. Rev. {\bf C55}, R2167 (1997).

\bibitem{rak96} S. Rakityansky {\it et al.},
  Phys. Rev. {\bf C53}, R2043 (1996).

\bibitem{sok94} G.A. Sokol {\it et al.}, Proc. of Int. Conf.
``Mesons and Nuclei at Intermediate Energies" May 3-7, Dubna,
Russia 1994, Eds.  M. Khankhasaev, Z. Kurmanov. Singapore 1995,
p.651.

\bibitem{sok91} G.A. Sokol, V.A. Tryasuchev,
  Kratkie Soobsh. Fiz. [English transl.: Sov. Phys. --
  Lebedev Institute Reports] {\bf 4}, 23 (1991) 23.

\bibitem{lvo98} A.I. L'vov, Proc. of the 7th Int. Conf. ``Mesons
and Light Nuclei", Pruhonice, Prague, Czech Republic, 31 Aug - 4
Sep 1998, Eds. J. Adam {\it et al.}, (World Scientific,
Singapore); nucl-th/9809054.

\bibitem{try95} A.I. Lebedev, V.A. Tryasuchev,
   Yad. Fiz. {\bf 58}, 642 (1995);
   V.A. Tryasuchev, private communication.

\end{thebibliography}
\end{document}